
\magnification=1200
\baselineskip=6mm
\nopagenumbers
\noindent
\null\vskip 1truecm
\centerline{\bf STATISTICS IN THE PROPOSITIONAL FORMULATION}
\centerline{\bf OF QUANTUM MECHANICS}
\vskip 2truecm
\centerline{D.R. Grigore\footnote*{e-mail: grigore@roifa.bitnet}}
\centerline{Department of Theoretical Physics, Institute of
Atomic Physics}
\centerline{Bucharest-Magurele,Romania}
\vskip 2truecm
\centerline{ABSTRACT}
\vskip 1truecm
We give a definition for the notion of statistics in the
lattice-theoretical (or propositional) formulation of quantum
mechanics of Birchoff, von Neumann and Piron. We show that this
formalism is compatible only with two types of statistics: Bose-Einstein
and Fermi-Dirac. Some comments are made about the connection between
this result and the existence of exotic statistics
(para-statistics, infinite statistics, braid statistics).
\vskip 1truecm
\vfil\eject

\footline={\hss\tenrm\folio\hss}
\pageno=1

{\bf 1. Introduction}

The lattice-theoretical formulation of quantum physics
(Birchoff-von Neumann-Piron) seems to be extremely well suited
for the treatement of many problems connected with the logical
foundations of a physical theory [1], [2]. The basic idea of this
formulation is that all elementary ("yes-no") statements which
can be made about a physical system can be organized in a
lattice structure
${\cal L}$.
For a pure quantum system the
corresponding lattice
${\cal L}$
is made up of all orthogonal
projectors in a given vector space of Hilbertian type
${\cal H}$:
${\cal L} = {\cal P}({\cal H})$.
A fundamental result asserts that this case is, essentially,
generic. Namely, the most general physical system is described
by a direct union of pure quantum lattices [1].

Among other things, the lattice-theoretical formulation of
quantum mechanics affords an answer to the question why two
(or many) quantum systems are usually described in the tensor
product Hilbert space of the individual Hilbert spaces of the
corresponding subsystems [3], [4]. This structure is a
consequence of the so-called "weak coupling" condition.
Essentially, this condition requires that the subsystems of the
composite system do not loose their individuality.
Mathematically, if
${\cal L}_{1},...,{\cal L}_{n}$
are the lattices of the individual subsystems and
${\cal L}_{0}$
is the lattice of the composite system, one requires the
existence of a map
$h: {\cal L}_{1}\times ...\times {\cal L}_{n} \rightarrow
{\cal L}_{0}$
with the following significance: if
$a_{1},...,a_{n}$
are properties of the subsystems
${\cal L}_{1},...,{\cal L}_{n}$
respectively, then
$h(a_{1},...,a_{n}) \in {\cal L}_{0}$
corresponds to the property: "the subsystem 1
has the property
$a_{1}$,...,
the subsystem $n$ has the property
$a_{n}$".
As we have said before, if
${\cal L}_{i} = {\cal L}({\cal H}_{i})$
with
$dim({\cal H}_{i}) \geq 3~~(i = 0,1,...,n)$
and the map $h$ has some reasonable properties, then one can
discover that in many cases of physical interest
${\cal H}_{0}$
has some tensorial nature.

An interesting problem is if this type of result can be extended
for systems of identical particles. An attempt in this
direction is announced in [5] where one finds the rather strange
result that for a system of identical particles only Fermi
statistics is allowed. We should note here that there are other
abstract definitions of the notion of statistics in the
framework of algebraic quantum theory [6].

The purpose of this paper is to give an alternative analysis for
systems of identical particles in the framework of the
lattice-theoretical formulation. We will give a reasonable "weak
coupling" condition for a system of identical particles and we
will be able to prove that, in quite general conditions, there
are only two possible statistics: Bose and Fermi.

The idea of the proof is suggested already by [3],  which uses
as an auxilliary result, a certain generalization of Wigner
theorem. So, the idea is to look for the "simplest" proof of
Wigner theorem and try to apply it to our situation. We have
found it profitable to use in such a way Uhlhorn proof of Wigner
theorem [7]. Using the idea of this proof (which will be briefly
presented) we will be able to give an alternative (and simpler in
our opinion) proof of the result of [3], [4] for the case of a
system composed of two different subsystems. In particular our
proof shows that the conditions impsed in [3], [4] on the map
$h: {\cal L}_{1}\times ...\times {\cal L}_{n} \rightarrow
{\cal L}_{0}$
can be relaxed. These topics are treated in Section 2.

In Section 3 we give our definition for a system of identical
particles and derive the result concerning the possible
statistics that was announced above. The proof will follow the
same lines as the one in Section  2. In the end of Section 3 we
will make some comments about the connection between our result
and the existence in the litterature of other types of
statistics as: parastatistics, infinite statistics [6] and braid
statistics [8], [9].
\vskip 1truecm
{\bf 2. Many-Particle Systems}

A. According to the lattice-theoretical philosophy one must
describe any physical system by a lattice
$({\cal L},<)$
which is complete, atomic, orthocomplemented, weakly modular and
satisfies the covering law. Usually we will omit the order
relation $<$ in writing the simbol of a lattice. Such a
lattice is also called a
{\it propositional system} [1] (see \$2.1). As regards to the
physical interpretation we mention only the following facts
without bothering about a precise mathematical formulation:

- the elements of
${\cal L}$
are interpreted as elementary ("yes-no") assertions about the system
(more precisely equivalence classes of "yes-no" questions)

- the order relation $<$ means logical implication

- the infimum operation
$\land$
has the meaning of the logical "AND"

- the orthocomplementation
${\cal L} \ni a \mapsto a' \in {\cal L}$
has the meaning of logical negation

- the atoms of
${\cal L}$
are interpreted as the states of the system

- the minimal element $0$ is interpreted as the property "the
system does not exist" and the maximal element $I$ is
interpreted as the property "the system exists".

The standard models of propositional systems are:

a) ${\cal P}(\Gamma)$;
here
$\Gamma$
is an arbitrary set and
${\cal P}(\Gamma)$
is the set of all subsets of
$\Gamma$.
The infimum is the intersection and the orthocomplementation is
the usual complementation. This is the {\it pure clasical} case.

b) ${\cal L}({\cal H})$;
here
${\cal H}$
is an arbitrary vector space of Hilbertian type and
${\cal L}({\cal H})$
is the set of all linear closed subspaces of
${\cal H}$.
The infimum is the intersection and the orthocomplementation
is the map associationg to any subspace of
${\cal H}$
its orthogonal suplement. This is the {\it pure quantum} case.

As asserted in [1] these cases are rather generic, in the sense
that the most general situation is obtained by taking a direct
union of pure quantum lattices; in this case the center of the
lattice
${\cal L}$
is a pure classical lattice $I$:
${\cal L} = \lor_{\alpha \in I} {\cal L}({\cal H}^{\alpha})$.
\vskip1 truecm

B. We schetch briefly the proof of Wigner theorem from [7]. The
idea of the proof will be afterwards adapted to the study of
many-particle systems. The purpose of Wigner theorem is to
classify symmetries of propositional systems. Such a symmetry is,
by definition, a structure-preserving map between propositional
systems. We adopt the following definition (which is equivalent
to other definitions in the litterature):

{\bf Definition 1:} Let
${\cal L}_{1}, {\cal L}_{2}$
be two propositional systems. A map
$h: {\cal L}_{1} \rightarrow {\cal L}_{2}$
is called a {\it symmetry} if it verifies:

(a) if
$p \in {\cal L}_{1}$
is an atom then
$h(p) \in {\cal L}_{2}$
is an atom

(b) for any
$a_{i} \in {\cal L}_{1}$
$i \in I$
an index set, one has:
$$
\land_{i \in I} h(a_{i}) = h(\land_{i \in I} a_{i})
$$

(c) $h(I_{1}) = I_{2}$

(d) for any
$a \in {\cal L}_{1}$
one has:
$$h(a)' = h(a')$$

{\bf Remark 1:} Usually (a) and (c) are replaced with the
condition of bijectivity.
\vskip0.5 truecm

Let us now suppose that the lattices
${\cal L}_{1}, {\cal L}_{2}$
are of the pure quantum type:
${\cal L}_{i} = {\cal L}({\cal H}_{i})$
where
${\cal H}_{i}$
is a vector space of Hilbertian type over the division ring
$D_{i}$
with
$dim({\cal H}_{i}) \geq 3~~(i = 1,2)$.
Then one can proceed to a rather exthaustive classification of
maps $h$ verifying Definition 1. We provide the main steps below.

1. If
$a_{1},...,a_{n} \in {\cal L}({\cal H}_{1})$
are atoms it is obvious what we mean when we say that
$a_{1},...,a_{n}$
are linear independent: for any
$x_{i} \in a_{i} \backslash \{0\}~(i = 1,...,n)$,
the vectors
$x_{1},...,x_{n} \in {\cal H}_{1}$
are linear independent. We also use the following notation: if
$x_{1} \in {\cal H}_{1} \backslash \{0\}$
then the atom containing
$x_{1}$
is denoted by
$D_{1}\cdot x_{1}$.

The first observation, following form Definition 1, is that
$a_{1},...,a_{n} \in {\cal L}({\cal H}_{1})$
are linear independent atoms {\it iff}
$h(a_{1}),...,h(a_{n}) \in {\cal L}({\cal H}_{2})$
are linear independent atoms.

2. We define now a map
$B: {\cal H}_{1} \rightarrow {\cal H}_{2}$
as follows:

- $B(0) = 0$

- for any
$x_{1} \in {\cal H}_{1} \backslash \{0\}$
we take
$B(x_{1})$
to be an arbitrary non-zero element in the atom
$h(D_{1}\cdot x_{1})$.

In this way we have
$$
D_{2}\cdot B(x_{1}) = h(D_{1}\cdot x_{1})
$$
for any
$x_{1} \in {\cal H}_{1} \backslash \{0\}$. It is clear that,
in general, the map $B$ will not be additive. However, one must
observe that there is a phase factor arbitrariness in the
definition of $B$. One takes advantage of this arbitrariness of $B$;
namely one shows that by apropriately modifying $B$ one can make
it an additive map.

3. Let
$x_{1}, y_{1} \in {\cal H}_{1}$,
be linear independent vectors. Using step 1 one easily establish
that one has:
$$
B(x_{1} + y_{1}) = \omega(x_{1},x_{1} + y_{1})~B(x_{1}) +
\omega(y_{1},x_{1} + y_{1})~B(y_{1})\eqno(2.1)
$$
where
$\omega: {\cal H}_{1} \times {\cal H}_{1} \rightarrow D_{2}
\backslash \{ 0\}$
is defined for the moment only for linear independent vectors.

4. Let now
$x_{1}, y_{1}, z_{1} \in {\cal H}_{1}$
be linear independent vectors. If one writes
$B(x_{1} + y_{1} + z_{1})$
in two different ways with the help of (2.1) one easily
discovers in this case a "cohomological" relationship:
$$
\omega(x_{1},y_{1})~\omega(y_{1},z_{1}) = \omega(x_{1},z_{1}).\eqno(2.2)
$$

5. An easy consequence of (2.2) is that for any
$x_{1}, y_{1} \in {\cal H}_{1}$
linear independent one has:
$$
\omega(x_{1},y_{1})~\omega(y_{1},x_{1}) = 1.\eqno(2.3)
$$

6. We now define
$\omega(x_{1},y_{1})$
for
$x_{1}, y_{1} \in {\cal H}_{1} \backslash  \{ 0\}$
linear dependent as follows. One takes
$z_{1} \in {\cal H}_{1}$
such that
$x_{1}$
and
$z_{1}$
are linear independent and tries to {\it define}
$\omega(x_{1},y_{1})$
by:
$$
\omega(x_{1},y_{1}) \equiv \omega(x_{1},z_{1})~\omega(z_{1},y_{1}).
\eqno(2.4)
$$

The right hand side does not depend on the choice of
$z_{1}$
above, so this definition is consistent.

7. Using the extension of $\omega$ defined above, one shows
easily that (2.3) is true for any
$x_{1}, y_{1} \in {\cal H}_{1} \backslash \{ 0\}$.

8. Next, one shows that (2.2) above is true for any
$x_{1}, y_{1}, z_{1} \in {\cal H}_{1} \backslash \{ 0\}$.
We note that it is at this point that one needs the restriction
$dim({\cal H}_{1}) \geq 3$.

9. Finally, one extends (2.1) for any
$x_{1}, y_{1} \in {\cal H}_{1} \backslash \{ 0\}$.

10. We are ready to redefine the map $B$ such that it becomes
additive. We take
$x_{1}^{0} \in {\cal H}_{1} \backslash \{ 0\}$
arbitrary but fixed and define for any
$x_{1} \in {\cal H}_{1} \backslash \{ 0\}$:
$$
\widetilde{B}(x_{1}) \equiv \omega(x_{1},x_{1}^{0})~B(x_{1}).\eqno(2.5)
$$

Then an easy computation shows that
$\widetilde{B}$
verifies for any
$x_{1}, y_{1} \in {\cal H}_{1} \backslash \{ 0\}$:
$$
\widetilde{B}(x_{1} + y_{1}) = \widetilde{B}(x_{1}) +
\widetilde{B}(y_{1}).\eqno(2.6)
$$

It is clear that (2.6) stays true even if
$x_{1}$
or
$y_{1}$
are zero. Because we still have
$D_{2}\cdot \widetilde{B}(x_{1}) = h(D_{1}\cdot x_{1})$
for any
$x_{1} \in {\cal H}_{1} \backslash \{ 0\}$
we might just well take instead of $B$ the new map
$\widetilde{B}$.

In conclusion, if Definition 1 is true, one can find an additive
map
$B: {\cal H}_{1} \rightarrow {\cal H}_{2}$:
$$
B(x_{1} + y_{1}) = B(x_{1}) + B(y_{1})~~~(\forall x_{1},y_{1}
\in {\cal H}_{1})\eqno(2.7)
$$
such that:
$$
D_{2}\cdot B(x_{1}) = h(D_{1}\cdot x_{1})~~~(\forall x_{1} \in
{\cal H}_{1} \backslash \{ 0\}).\eqno(2.8)
$$

11. It is not difficult to show that $B$ also verifies:
$$
Im(B) = {\cal H}_{2}.\eqno(2.9)
$$

It follows that $B$ is a bijective map.

12. Now it is rather easy to prove that there exists a map
$\varphi: D_{1} \rightarrow D_{2}$
such that:
$$
B(\lambda_{1} x_{1}) = \varphi(\lambda_{1})~B(x_{1})~~~
(\forall \lambda_{1} \in D_{1}, \forall x_{1} \in {\cal H}_{1}).
\eqno(2.10)
$$

Moreover, the map
$\varphi$
verifies for any
$\lambda_{1}, \mu_{1} \in D_{1}$:
$$
\varphi(\lambda_{1} + \mu_{1}) = \varphi(\lambda_{1}) +
\varphi(\mu_{1})\eqno(2.11)
$$
$$
\varphi(\lambda_{1}\mu_{1}) = \varphi(\lambda_{1})~
\varphi(\mu_{1}) \eqno(2.12).
$$

It is not hard to convince oneself that
$\varphi$
is indeed a division ring isomorphism. So in fact one can take
$D_{1} = D_{2} (\equiv D_{0})$.

13. From Definition 1 it follows that the map $h$ preserves the
orthogonality relationship. If the division ring
$D_{0}$
is commutative, it follows that
$\varphi$
also verifies:
$$
\bar{\varphi}(\lambda) = \varphi(\bar{\lambda})\eqno(2.13)
$$
where
$\lambda \rightarrow \bar{\lambda}$
is the involution of
$D_{0}$.

{\bf Remark 2:} The usual cases
$D_{0} = R, C, H$
can be analysed in detail as in [2]. If
$D_{0} = R, H$
one finds that one can take
$\varphi = id$
so $B$ becomes a linear map, and if
$D_{0} = C$
one has two cases:
$\varphi(\lambda) = \lambda$
and
$\varphi(\lambda) = \bar{\lambda}$
corresponding to $B$ linear and respectively antilinear. We have
recovered the usual statement of Wigner theorem.
\vskip0.5 truecm
{\bf Remark 3:} One can also show that there exists
$\delta \in D_{0} \backslash \{ 0\}$
such that
$\forall x_{1}, y_{1} \in {\cal H}_{1}$ one has:
$$
<B(x_{1}),B(y_{1})>_{{\cal H}_{2}} = \delta
\varphi(<x_{1},y_{1}>_{{\cal H}_{1}})\eqno(2.14)
$$
with
$\bar{\delta} = \delta$.
The idea is to adopt (2.14) as the definition for
$\delta$
as a function of
$x_{1}, y_{1}, x_{2}, y_{2}$
and to show that in fact it is a constant.

If
${\cal H}_{1}$
and
${\cal H}_{2}$
are Hilbert spaces over
$R,C$
or
$H$
one can show that in fact
$\delta > 0$.

So, by a rescaling, $B$ can be made an isometry.
\vskip1 truecm

C. Now we come to the study of a composite system. We formulate
[3], [4]:

{\bf Definition 2:} Let
$\{ {\cal L}_{i}\}_{i=0}^{3}$
be three propositional systems with
${\cal L}_{1} \not= {\cal L}_{2}$.
We say that the system
${\cal L}_{0}$
is {\it composed of the subsystems}
${\cal L}_{1}$
{\it and}
${\cal L}_{2}$
if there exists a map
$h:{\cal L}_{1} \times {\cal L}_{2} \rightarrow {\cal L}_{0}$
verifying:

(a) if
$p_{1} \in {\cal L}_{1}$
and
$p_{2} \in {\cal L}_{2}$
are atoms then
$h(p_{1},p_{2}) \in {\cal L}_{0}$
is an atom

(b) $\forall a_{i} \in {\cal L}_{1}~(i \in I)$
and
$\forall b_{i} \in {\cal L}_{2}(i \in I)$~
($I$ is an index set), one has:
$$
\land_{i \in I} h(a_{i},b_{i}) = h(\land_{i \in I} a_{i},
\land_{i \in I} b_{i})
$$

(c) $h(I_{1},I_{2}) = I_{0}$

(d) $\forall a \in {\cal L}_{1}$
and
$\forall b \in {\cal L}_{2}$
one has:
$$
h(a,b)' = h(a',b')
$$

{\bf Remark 4:} The existence of the map $h$ can be also called
the {\it weak coupling condition} [3]. Indeed, when postulating
the existence of such a map one implicitely asumes that the subsystems
${\cal L}_{1}$
and
${\cal L}_{2}$
do not loose their individuality. It is clear that the proposition
$h(a_{1},a_{2})$
coresponds to the property: "the subsystem 1 has the property
$a_{1}$
and the subsystem 2 has the property
$a_{2}$".
\vskip0.5 truecm

{\bf Remark 5:} The physical interpretation of the axioms
(a)-(d) above is rather transparent [3], [4]. We note however
that we did not include in this definition the condition:
$h(a_{1},I_{2}) \leftrightarrow h(I_{1},a_{2})~~\forall a_{1} \in
{\cal L}_{1},~~\forall a_{2} \in {\cal L}_{2}$
which is explicitely admitted in [3], [4]. In fact, the analysis
below will show that this condition is redundant.
\vskip0.5 truecm

We proceed now with the classification of maps $h$ in the case when
${\cal L}_{i} = {\cal L}({\cal H}_{i})$
where
${\cal H}_{i}$
is a vector space of Hilbertian type over the division ring
$D_{i}$~(i = 0, 1, 2).

We also admit that
$dim({\cal H}_{i}) \geq 3~~~(i =0,1,2)$.
We will follow closedly the steps 1-12 of part B.

1. One can show rather easily that the atoms
$a_{i} \in {\cal L}_{1}~~(i \in I)$
are linear independent {\it and} the atoms
$b_{j} \in {\cal L}_{2}~~(j \in J)$
are linear independent {\it iff} the atoms
$h(a_{i},b_{j}) \in {\cal L}_{0}~~(i \in I,j \in J)$
are linear independent.

2. We now define a map
$B: {\cal H}_{1} \times {\cal H}_{2} \rightarrow {\cal H}_{0}$
as follows:

- $B(0,x_{2}) = 0,~~~B(x_{1},0) = 0~~(\forall x_{1} \in {\cal H}_{1},
\forall x_{2} \in {\cal H}_{2})$

- for $x_{1} \in {\cal H}_{1} \backslash \{ 0\},
x_{2} \in {\cal H}_{2} \backslash \{ 0\}$
we take
$B(x_{1},x_{2})$
to be an arbitrary non-zero vector in the atom
$h(D_{1}\cdot x_{1},D_{2}\cdot x_{2})$.
In this way we have
$D_{0}\cdot B(x_{1},x_{2}) = h(D_{1}\cdot x_{1},D_{2}\cdot x_{2})$.
Like in part B we will use the phase arbitrariness in the
definition of $B$ to make it biadditive.

3. We fix the vector
$x_{2} \in {\cal H}_{2} \backslash \{ 0\}$.
Then we can repeat steps A3-A10 for the map
$x_{1} \rightarrow B(x_{1},x_{2})$
and we succeed to redefine $B$ such that it is additive in the
first argument:
$$
B(x_{1} + y_{1},x_{2}) = B(x_{1},x_{2}) + B(y_{1},x_{2})
{}~~\forall x_{1},y_{1} \in {\cal H}_{1},~\forall x_{2} \in {\cal H}_{2}
\eqno(2.15)
$$

4. Now we fix
$x_{1} \in {\cal H}_{1} \backslash \{ 0\}$
and repeat steps A3-A9 for the map
$x_{2} \rightarrow B(x_{1},x_{2})$.
We find that there exists a map:
$\omega_{x_{1}}: ({\cal H}_{2} \backslash \{ 0\}) \times
({\cal H}_{2} \backslash \{ 0\}) \rightarrow D_{0}$
such that
$\forall x_{2},y_{2} \in {\cal H}_{2} \backslash \{ 0\}$:
$$
B(x_{1},x_{2} + y_{2}) =
\omega_{x_{1}}(x_{2},x_{2} + y_{2})~B(x_{1},x_{2}) +
\omega_{x_{1}}(y_{2},x_{2} + y_{2})~B(x_{1},y_{2}).\eqno(2.16)
$$

The function
$\omega_{x_{1}}$
also verifies
$\forall x_{2},y_{2},z_{2} \in {\cal H}_{2} \backslash \{ 0\}$:
$$
\omega_{x_{1}}(x_{2},y_{2})~\omega_{x_{1}}(y_{2},z_{2}) =
\omega_{x_{1}}(x_{2},z_{2})\eqno(2.17)
$$
and:
$$
\omega_{x_{1}}(x_{2},y_{2})~\omega_{x_{1}}(y_{2},x_{2}) = 1.\eqno(2.18)
$$

5. We want to apply the trick A10 to get additivity in the
second argument of $B$ without ruining the same property in the
first argument. For this we have to show first that in fact
$\omega_{x_{1}}(\cdot,\cdot)$
does not depend on
$x_{1}$.
This is rather simple. One takes
$x_{i},y_{i} \in {\cal H}_{i}~(i = 1,2)$
linear independent and makes in (2.16)
$x_{1} \rightarrow x_{1} + y_{1}$.
If we use now the additivity (2.15) we arrive quite easily at the
following relationship:
$$
\omega_{x_{1}}(x_{2},y_{2})  = \omega_{y_{1}}(x_{2},y_{2}).\eqno(2.19)
$$

If
$x_{1},y_{1} \in {\cal H}_{1} \backslash \{ 0\}$
verify
$D_{1}\cdot x_{1} = D_{1}\cdot y_{1}$
one takes
$z_{1} \in {\cal H}_{1} \backslash \{ 0\}$
such that
$D_{1}\cdot x_{1} \not= D_{1}\cdot z_{1}$
and has from (2.19):
$$
\omega_{x_{1}}(x_{2},y_{2})  = \omega_{z_{1}}(x_{2},y_{2}) =
\omega_{y_{1}}(x_{2},y_{2})
$$

In this way one extends (2.19) for all
$x_{1},y_{1} \in {\cal H}_{1} \backslash \{ 0\}$.
We have only the restriction
$D_{2}\cdot x_{2} \not= D_{2}\cdot y_{2}$.
But if
$D_{2}\cdot x_{2} = D_{2}\cdot y_{2}$,
then
$\omega_{x_{1}}(x_{2},y_{2})$
is defined according to A6 as follows:
$$
\omega_{x_{1}}(x_{2},y_{2})  = \omega_{x_{1}}(x_{2},z_{2})~
\omega_{x_{1}}(z_{2},y_{2})
$$
where
$z_{2} \in {\cal H}_{2} \backslash \{ 0\}$
verifies
$D_{2}\cdot x_{2} \not= D_{2}\cdot y_{2}$. (see (2.4)).
The right hand side of this relation does not depend on
$x_{1}$
according to what has been proved so far . So we have succeeded
to extend (2.19) to all
$x_{i},y_{i} \in {\cal H}_{i} \backslash \{ 0\}~(i = 1,2)$.

6. It follows that in fact in (2.16)-(2.18)
$\omega_{x_{1}}(\cdot,\cdot)$
does not depend on
$x_{1}$
so we have for any
$x_{2},y_{2} \in {\cal H}_{2} \backslash \{ 0\}$:
$$
B(x_{1},x_{2} + y_{2}) =
\omega(x_{2},x_{2} + y_{2})~B(x_{1},x_{2}) +
\omega(y_{2},x_{2} + y_{2})~B(x_{1},y_{2}).\eqno(2.20)
$$
where the  function
$\omega$
also verifies
$$
\omega(x_{2},y_{2})~\omega(y_{2},z_{2}) =
\omega(x_{2},z_{2})\eqno(2.21)
$$
and:
$$
\omega(x_{2},y_{2})~\omega(y_{2},x_{2}) = 1.\eqno(2.22)
$$

Now we apply the trick A10 for the map
$x_{2} \mapsto B(x_{1},x_{2})$
and we succeed to make it additive in the second argument,
preserving in the mean time the same property in the first
argument. So, beside (2.15) we have:
$$
B(x_{1},x_{2} + y_{2}) = B(x_{1},x_{2}) + B(x_{1},y_{2})\eqno(2.23)
$$
and:
$$
D_{0}\cdot B(x_{1},x_{2}) = h(D_{1}\cdot x_{1},D_{2}\cdot
x_{2}).\eqno(2.24)
$$

7. Like in part A it is not difficult to show that B also verifies:
$$
Im(B) = {\cal H}_{0}.\eqno(2.25)
$$

8. Step A12 goes now practically unchanged. One proves the
existence of a map
$\varphi: D_{1} \times D_{2} \rightarrow D_{0}$
such that:
$$
B(\lambda_{1} x_{1}, \lambda_{2} x_{2}) = \varphi(\lambda_{1},
\lambda_{2})~B(x_{1},x_{2}).\eqno(2.26)
$$

Moreover, the map
$\varphi$
verifies:
$$
\varphi(\lambda_{1} + \mu_{1},\lambda_{2} + \mu_{2}) =
\varphi(\lambda_{1},\lambda_{2}) +
\varphi(\lambda_{1},\mu_{2}) +
\varphi(\mu_{1},\lambda_{2}) +
\varphi(\mu_{1},\mu_{2})\eqno(2.27)
$$
$$
\varphi(\lambda_{1}  \mu_{1},\lambda_{2}  \mu_{2}) =
\varphi(\lambda_{1},\lambda_{2})~
\varphi(\mu_{1},\mu_{2}).\eqno(2.28)
$$

9. Like at A13 one can show that if
$D_{0}$
is commutative
$\varphi$
also satisfies:
$$
\bar{\varphi}(\lambda_{1},\lambda_{2}) =
\varphi(\bar{\lambda}_{1},\bar{\lambda}_{2})\eqno(2.29)
$$
where the bar denotes the corresponding involutions of
$D_{i}~~(i = 1,2)$.

{\bf Remark 6:} If
$D_{0} = D_{1} = D_{2} = R$
we have
$\varphi(\lambda_{1},\lambda_{2}) = \lambda_{1}\lambda_{2}$
and if
$D_{0} = D_{1} = D_{2} = C$
we have four possibilities:
$$
\varphi(\lambda_{1},\lambda_{2}) = \lambda_{1}\lambda_{2},~~~
\varphi(\lambda_{1},\lambda_{2}) = \bar{\lambda}_{1}\bar{\lambda}_{2}
$$
$$
\varphi(\lambda_{1},\lambda_{2}) = \bar{\lambda}_{1}\lambda_{2},~~~
\varphi(\lambda_{1},\lambda_{2}) = \lambda_{1}\bar{\lambda}_{2}.
$$

On the contrary if
$D_{0} = D_{1} = D_{2} = H$
one can prove that (2.27)+(2.28) have no solution. It is quite
possible that, in general, the weak coupling problem does not
admit solutions if
$D_{1}$ and
$D_{2}$
are non-commutative division rings. (See however in connection
with this problem [10].)
\vskip0.5 truecm

10. Taking into account the Remark above it is intersting to
consider the particular case when
$D_{0} = D_{1} = D_{2}$
is a {\it commutative} division ring and
$\varphi: D_{0} \times D_{0} \rightarrow D_{0}$ is:
$$
\varphi(\lambda_{1},\lambda_{2}) = \lambda_{1}\lambda_{2}.\eqno(2.30)
$$

We have:

{\bf Theorem 1:} In the conditions above one can take
${\cal H}_{0} = {\cal H}_{1} \otimes {\cal H}_{2}$.
Moreover, if
${\cal H}_{i}$
are Hilbert spaces and we identify
${\cal L}({\cal H}_{i})$
with the lattice of the orthogonal projectors in
${\cal H}_{i}$,
then the map $h$ is:
$$
h(a_{1},a_{2}) = a_{1} \otimes a_{2}.\eqno(2.31)
$$

{\bf Proof:} One has to check that the map $B$ has the
universality properties defining the tensor product (see e.g.
[11]).

$\otimes_{1}$:
We have
$Im(B) = {\cal H}_{0}$
according to step 7.

$\otimes_{2}$:
Let
${\cal H}$
be vector space over the division ring
$D_{0}$
and
$g: {\cal H}_{1} \times {\cal H}_{2} \rightarrow {\cal H}$
be a bilinear map. Our purpose is to identify a linear map
$f: {\cal H}_{0} \rightarrow {\cal H}$
such that:
$$
f \circ B = g.\eqno(2.32)
$$

We procced as follows. First, let us consider
$x_{i},y_{i} \in {\cal H}_{i}~(i = 1,2)$
such:
$$
B(x_{1},x_{2}) = B(y_{1},y_{2}).\eqno(2.33)
$$

Then, one easily establish that
$y_{i} = \lambda_{i} x_{i}~(i = 1,2)$
with
$\lambda_{1}, \lambda_{2} \in D_{0}$
verifying
$\lambda_{1} \lambda_{2} = 1$.
It follows that
$$
g(y_{1},y_{2}) = g(x_{1},x_{2}).\eqno(2.34)
$$

So, taking into account
$\otimes_{1}$
we define
$f: {\cal H}_{0} \rightarrow {\cal H}$
as follows:

- for elements of the form
$B(x_{1},x_{2})$:
$$
f(B(x_{1},x_{2})) = g(x_{1},x_{2})\eqno(2.35)
$$
(and this is consistent because of the implication (2.33)
$\Rightarrow$ (2.34)).

- for all the other elements we extend $f$ by continuity:
$$
f(\sum_{i=1}^{n} B(x_{i},y_{i})) = \sum_{i=1}^{n}
g(x_{i},y_{i})\eqno(2.36)
$$
(and again one can prove the consistency of this definition).
But (2.35) and (2.36) gives (2.32). Q. E. D.

{\bf Remark 7:} The theorem above takes care of the case
$D_{0} = R$.
In the case
$D_{0} = C$
we have the four possibilities from Remark 6. ( We note that
in this case
${\cal H}_{i}$
are Hilbert spaces according to Amemiya-Araki theorem [1], [2]).
We still have to analyse the last three of them. We define the
antilinear maps
$\alpha_{i}: {\cal H}_{i} \rightarrow ({\cal H}_{i})^{*}~(i =
1,2)$
by:
$$
<\alpha_{i}(x_{i}),y_{i}> = (x_{i},y_{i})_{{\cal H}_{i}}.\eqno(2.37)
$$

Here
$( , )_{{\cal H}_{i}}$
is the scalar product on
${\cal H}_{i}$
and
$< , >$
is the duality form between
${\cal H}_{i}$
and
$({\cal H}_{i})^{*}$.
For the case
$\varphi(\lambda_{1},\lambda_{2}) =
\bar{\lambda}_{1}\bar{\lambda}_{2}$
we define:
$B_{12}: ({\cal H}_{1})^{*} \times ({\cal H}_{2})^{*} \rightarrow
{\cal H}_{0}$ by:
$$
B_{12}(x_{1},x_{2}) = B(\alpha_{1}^{-1}(x_{1}),\alpha_{2}^{-1}(x_{2}))
$$
and note that
$B_{12}$
is bilinear so applying the theorem above we can take
${\cal H}_{0} = ({\cal H}_{1})^{*} \times ({\cal H}_{2})^{*}
\cong {\cal H}_{1} \times {\cal H}_{2}$.
In the last two cases one procceeds similarly and find out that
it is possible to take
${\cal H}_{0} = ({\cal H}_{1})^{*} \times {\cal H}_{2}
\cong {\cal H}_{1} \times ({\cal H}_{2})^{*}$.
\vskip0.5 truecm

{\bf Remark 8:} A formula of the type (2.14) can be also proved
in this case, namely:
$$
<B(x_{1},x_{2}),B(y_{1},y_{2})>_{{\cal H}_{0}} = \delta
\varphi (<x_{1},y_{1}>_{{\cal H}_{1}},<x_{2},y_{2}>_{{\cal H}_{2}})\eqno(2.38)
$$
for some
$\delta \in D_{1} \backslash \{0\}$.
\vskip0.5 truecm

{\bf Remark 9:} One can extend the results obtained up till now
for the more general case when
${\cal L}_{0}, {\cal L}_{1}$
and
${\cal L}_{2}$
are systems with superselection rules i.e.
${\cal L}_{i} = \lor_{\alpha_{i} \in I_{i}}
{\cal L}({\cal H}_{i}^{(\alpha_{i})})~(i = 0,1,2)$.
Here
$I_{0}, I_{1}$
and
$I_{2}$
are some index sets. We proceed in analogy to [1] \$3-2. First
we note that the map $h$ preserves the relationship of
compatibility, namely if
$a_{1}$
is in the center of
${\cal L}_{1}$
and
$a_{2}$
is in the center of
${\cal L}_{2}$,
then
$h(a_{1},a_{2})$
is in the center of
${\cal L}_{0}$.
It follows that the map $h$ induces a map
$\hat{h}: {\cal P}(I_{1}) \times {\cal P}(I_{2}) \rightarrow
{\cal P}(I_{0})$
where
${\cal P}(I_{i})$
are classical propositional systems (see the begining of part
A). The map
$\hat{h}$
also verifies the axioms of Definition 1. In this case one can
easily discover that
$I_{0} \cong I_{1} \times I_{2}$
[12] so in fact we have
${\cal L}_{0} \cong \lor_{\alpha_{1} \in I_{1},\alpha_{2} \in I_{2}}
{\cal L}({\cal H}_{0}^{(\alpha_{1},\alpha_{2})})$
and
$h({\cal L}({\cal H}_{1}^{(\alpha_{1})}),{\cal L}({\cal H}_{2}^{(\alpha_{2})}))
= {\cal L}({\cal H}_{0}^{(\alpha_{1}\alpha_{2})})~~\forall
\alpha_{1} \in I_{1}, \forall \alpha_{2} \in I_{2}$.
It is now clear that for all the maps
$h^{\alpha_{1}\alpha_{2}} \equiv h\vert_{{\cal L}({\cal H}_{1}^{(\alpha_{1})})
\times {\cal L}({\cal H}_{2}^{(\alpha_{2})})}$
one can apply the previous analysis.
\vskip0.5 truecm

{\bf Remark 10:} It is obvious that the analysis contained in
this Section can be easily extended to the case when the system
${\cal L}_{0}$
is composed of more that two subsystems
${\cal L}_{1},...,{\cal L}_{n}~(n > 2)$.
This kind of generalization seems to be more cumbersome to do if
one adopts the line of argument in [3], [4].
\vskip1 truecm

{\bf 3. Systems of Identical Particles}

A. We try here to propose a definition for a system composed of
two {\it identical} subsystems by modifying as little as
possible Definition 1. It is clear from the begining that one
must take
${\cal L}_{1} = {\cal L}_{2}$
i.e. we have a map
$h:{\cal L}_{1} \times {\cal L}_{1} \rightarrow {\cal L}_{0}$.
Also, we expect that, because of the identity of the the
subsystems, this map is symmetric. The physical interpetation of
$h$ must be the following one: if
$a,b \in {\cal L}_{1}$
are two properties then
$h(a,b)$
correspond to the property: "one of the subsystem is in the
state $a$ and the other is in the state $b$". Because of this
(natural) interpretation it follows that we cannot expect that
item (b) of Definition 1 holds in this case. If we could
interpret the supremum operation
$\lor$
as the logical "OR" we would be tempted to substitute (b) of
Definition 1 by something of the type:
$$
h(a,b) \land h(c,d) = h(a \land c,b \land d) \lor h(a \land d,b
\land c).
$$

However, it is known that
$\lor$
can be interpreted as the logical "OR" only in the pure
classical case. (Indeed, in the pure quantum case one can easily
find two properties
$a,b \in {\cal L}({\cal H})$
such that the logical proposition
$aORb$
corresponds to no element in
${\cal L}({\cal H})$).
So, the relation above cannot hold.

After this discussion we make an attempt for a convenient definition.

{\bf Definition 2:} We say that the propositional system
${\cal L}_{0}$
is {\it composed of two identical subsystems} if there exists a
propositional system
${\cal L}_{1}$
and a map
$h:{\cal L}_{1} \times {\cal L}_{1} \rightarrow {\cal L}_{0}$
verifying:

($a_{1}$) if
$p_{1}, p_{2} \in {\cal L}_{1},~p_{1} \not= p_{2}$
are atoms, then
$h(p_{1},p_{2}) \in {\cal L}_{0}$
is an atom

($a_{2}$) if
$p \in {\cal L}_{1}$
is an atom then
$h(p,p) \in {\cal L}_{0}$
is either $0$ or a atom

($a_{3}$) if
$a,b,c,d \in {\cal L}_{1}$
are atoms and
$$
h(a,b) = h(c,d) \not= 0
$$
then
$a = c,~b = d$
or
$a = d,~b = c$.

($b_{1}$) if
$a < c,d$
and
$b < c,d$,
then
$h(a,b) < h(c,d)$

($b_{2}$) the atoms
$a_{i} \in {\cal L}_{1}~(i \in I)$
are linear independent {\it iff} all the distinct non-zero atoms
of the form
$h(a_{i},a_{j})$
are linear independent

($c_{1}$)
$
\lor_{p_{1},p_{2}  = atoms} h(p_{1},p_{2}) = I_{0}
$

($c_{2}$)
$
h(I_{1},I_{2}) = I_{0}
$

(d)
$h(a,b)' = h(a',b')
$

(e)
$
h(a,b) = h(b,a)
$.

{\bf Remark 11:} All the axioms exept
$(b_{2}$)
and
$(c_{1})$
are rather easy to interpret from the physical point of view.
For instance,
$(a_{2})$
takes into account the logical possibility that the two
identical system cannot be in the same state.
\vskip0.5 truecm

Because
$(b_{2})$
and
$(c_{1})$
have a certain degree of naturalness we think it is interesting
to analyse in detail the consequence of this definition.

We proceed now with this analysis on the lines of Subsection 2C
in the case
${\cal L}_{i} = {\cal L}({\cal H}_{i})$
where
${\cal H}_{i}$
is a vector space of Hilbertian type over the division ring
$D_{i}~(i = 0,1)$
and
$dim({\cal H}_{1}) \geq 4$.

1. We define a map
$B: {\cal H}_{1} \times {\cal H}_{1} \rightarrow {\cal H}_{0}$
as follows:

- $B(0,x_{2}) = B(x_{1},0) = 0,~\forall x_{1}, x_{2} \in
{\cal H}_{1}$

- if
$x_{1} \in {\cal H}_{1} \backslash \{ 0\}$
and
$h(D_{1}\cdot x_{1},D_{1}\cdot x_{1}) = 0$
then
$B(x_{1},x_{1}) = 0$

- if $x_{1},x_{2} \in {\cal H}_{1} \backslash \{ 0\}$
and
$x_{1} \not= x_{2}$
or if
$x_{1} = x_{2}$
but
$h(D_{1}\cdot x_{1},D_{1}\cdot x_{1}) \not= 0$,
then
$B(x_{1},x_{2})$
is an arbitrary non-zero element of the atom
$h(D_{1}\cdot x_{1},D_{1}\cdot x_{2})$

It is clear that we have
$D_{0}\cdot B(x_{1},x_{2}) = h(D_{1}\cdot x_{1},D_{1}\cdot x_{2})$.

2. From Definition 2 $(b_{2})$
one can show that for any
$x_{2} \in {\cal H}_{1} \backslash \{ 0\}$
there exists a function
$\omega_{x_{2}}: ({\cal H}_{1} \backslash \{ 0\})
\times ({\cal H}_{1} \backslash \{ 0\}) \rightarrow D_{0}$
such that
$\forall x_{1}, y_{1}, x_{2}$
linear independent we have:
$$
B(x_{1} + y_{1},x_{2}) = \omega_{x_{2}}(x_{1},x_{1} + y_{1})~
B(x_{1},x_{2}) + \omega_{x_{2}}(y_{1},x_{1} + y_{1})~
B(y_{1},x_{2}).\eqno(3.1)
$$

3. Next, one takes
$x_{1}, y_{1}, z_{1}, x_{2} \in {\cal H}_{1}$
linear independent and shows that we have:
$$
\omega_{x_{2}}(x_{1},y_{1})~\omega_{x_{2}}(y_{1},z_{1}) =
\omega_{x_{2}}(x_{1},z_{1})\eqno(3.2)
$$
and
$$
\omega_{x_{2}}(x_{1},y_{1})~\omega_{x_{2}}(y_{1},x_{1}) = 1\eqno(3.3)
$$

4. Like at 2A, we now extend the function
$\omega_{x_{2}}$
to other values of
$x_{1}, y_{1}$.
One must use definitions of the type (2.4) and prove their
consistency. It is at this step that one needs the condition
$dim({\cal H}_{1}) \geq 4$.

5. Using the definition of
$\omega$
from above, one extends (3.1)-(3.3) to all
$x_{1}, y_{1}, z_{1}, x_{2} \in {\cal H}_{1} \backslash \{ 0\}$.
So, applying the trick 2A.10 one succeds to make the map $B$
additive in the first argument:
$$
B(x_{1} + y_{1},x_{2}) = B(x_{1},x_{2}) + B(y_{1},x_{2}).\eqno(3.4)
$$

Next, we apply the trick 2C.5-6 and obtain additivity in the
second argument also:
$$
B(x_{1},x_{2} + y_{2}) = B(x_{1},x_{2}) + B(x_{1},y_{2}).\eqno(3.5)
$$

Moreover, we still have:
$$
D_{0}\cdot B(x_{1},x_{2}) = h(D_{1}\cdot x_{1},D_{1}\cdot
x_{2}).\eqno(3.6)
$$

Of course, in obtaining (3.4) and (3.5) there are more cases to
study (by comparison with Subsection 2C) because there are more
possibilities of linear dependence which has to be studied case
by case. Exept for tediousness, the proof is not very difficult.
\vskip0.5 truecm

{\bf Remark 12:} One can extend the result above for a system of
$n$ identical subsystems
$(n > 2)$
if one takes
$dim({\cal H}_{1})$
sufficiently large.

6. From Definition 2 $(c_{1})$ one gets immediately:
$$
Im(B) = {\cal H}_{0}.\eqno(3.7)
$$

7. Now we come to the most interesting part, namely we use
Definition 2 (e) which expresses the identity of the subsystems.
This will impose some additional restrictions on the map $B$.

Indeed, Fron Definition 2 (e) it follows the existence of a map
$\varepsilon: {\cal H}_{1} \times {\cal H}_{1} \rightarrow D_{0}$
such that:
$$
B(x_{1},x_{2}) = \varepsilon(x_{1},x_{2}) B(x_{2},x_{1}).\eqno(3.8)
$$

It is an easy matter now to use the biadditivity of $B$ and
prove that in fact
$\varepsilon$
is a constant element of
$D_{0}$:
$$
B(x_{1},x_{2}) = \varepsilon B(x_{2},x_{1}).\eqno(3.9)
$$

It is clear that
$\varepsilon$
is constrained by:
$$\varepsilon^{2} = 1~~\Leftrightarrow (\varepsilon + 1)
(\varepsilon - 1) = 0.
$$

So, if we suppose that the division ring
$D_{0}$ does not admit divisors of $0$, then we have exactly two
possibilities:
$\varepsilon = 1$
and
$\varepsilon = -1$.
We call these possibilities {\it statistics}. When we have:
$$
B(x_{2},x_{1}) = B(x_{1},x_{2})\eqno(3.10)
$$
we say that we have {\it Bose-Einstein statistics} and when we have:
$$
B(x_{2},x_{1}) = -B(x_{1},x_{2})\eqno(3.11)
$$
we say that we have {\it Fermi-Dirac statistics}.
\vskip0.5 truecm

{\bf Remark 13:} The result above can be extended to the case of
$n$ identical subsystems (for $n > 2$). Indeed, if we have
already the additivity of the map
$B:{\cal H}_{1}^{\times n} \rightarrow
{\cal H}_{0}$
in all arguments, then instead of (3.9) we will find:
$$
B(x_{\sigma(1)},...,x_{\sigma(n)}) = \varepsilon(\sigma)
B(x_{1},...,x_{n})\eqno(3.12)
$$
for any permutation
$\sigma \in {\cal P}_{n}$
of the numbers
$1,...,n$.
{}From (3.12) it follows that
$\varepsilon: {\cal P}_{n} \rightarrow D_{0}$
is an one-dimensional representation of the permutation group
${\cal P}_{n}$.
It is known that (at least in the case
$D_{0} = R,C$)
one has exactly two such representations:
$\varepsilon = id$
and
$\varepsilon(\sigma) = sign(\sigma)$.

8. Like at 2C.8 one finds out that there exists a map
$\varphi: D_{1} \times D_{1} \rightarrow D_{0}$
such that:
$$
B(\lambda_{1} x_{1},\lambda_{2} x_{2}) =
\varphi(\lambda_{1},\lambda_{2})~B(x_{1},x_{2}).\eqno(3.13)
$$

This map verifies:
$$
\varphi(\lambda_{1} + \mu_{1},\lambda_{2} + \mu_{2}) =
\varphi(\lambda_{1},\lambda_{2}) +
\varphi(\lambda_{1},\mu_{2}) +
\varphi(\mu_{1},\lambda_{2}) +
\varphi(\mu_{1},\mu_{2})\eqno(3.14)
$$
$$
\varphi(\lambda_{1}  \mu_{1},\lambda_{2}  \mu_{2}) =
\varphi(\lambda_{1},\lambda_{2})~
\varphi(\mu_{1},\mu_{2}).\eqno(3.15)
$$
$$
\varphi(\lambda_{2},\lambda_{1}) =
\varphi(\lambda_{1},\lambda_{2}).\eqno(3.16)
$$

If
$D_{0}$
is a commutative division ring
$\varphi$
also satisfies:
$$
\bar{\varphi}(\lambda_{1},\lambda_{2}) =
\varphi(\bar{\lambda}_{1},\bar{\lambda}_{2})\eqno(3.17)
$$

For
$D_{0} = R$
we have
$\varphi(\lambda_{1},\lambda_{2}) = \lambda_{1} \lambda_{2}$
and for
$D_{0} = C$
we have another possibility, namely
$\varphi(\lambda_{1},\lambda_{2}) = \bar{\lambda}_{1}
\bar{\lambda}_{2}$.

9. Like in the precceding Section it is interesting to consider
the particular case when
$D_{0} = D_{1}$
is a commutative division ring and
$\varphi: D_{0} \times D_{0} \rightarrow D_{0}$
is:
$$
\varphi(\lambda_{1},\lambda_{2}) = \lambda_{1} \lambda_{2}.\eqno(3.18)
$$

We have in analogy to Theorem 1:

{\bf Theorem 2:} In the conditions above one can take
${\cal H}_{0} = \lor^{2}{\cal H}_{1}$
if $B$ is symmetric and
${\cal H}_{0} = \land^{2}{\cal H}_{1}$
if $B$ is antisymmetric.

{\bf Remark 14:} We see that the structure of symmetric and
antisymmetric tensor product emerges naturally in our framework.
In particular, this is the case when
$D_{0} = R,C$.
\vskip0.5 truecm

{\bf Remark 15:} One can also prove a formula of the type (2.38):
$$
<B(x_{1},x_{2}),B(y_{1},y_{2})>_{{\cal H}_{0}} =
$$
$$
{1\over 2} \delta
[\varphi(<x_{1},y_{1}>_{{\cal H}_{1}},<x_{2},y_{2}>_{{\cal H}_{1}})
\pm \varphi(<x_{1},y_{2}>_{{\cal H}_{1}},<x_{2},y_{1}>_{{\cal H}_{1}})]
\eqno(3.19)
$$
with
$+(-)$
if $B$ is symmetric (antisymmetric).

For the case of $n$ identical subsystems, the coresponding
formulae are:
$$
<B(x_{1},...,x_{n}),B(y_{1},...,y_{n})>_{{\cal H}_{0}} =
$$
$$
{1\over n!} \delta \sum_{\sigma \in {\cal P}_{n}}
\varphi(<x_{1},y_{\sigma(1)}>_{{\cal H}_{1}},...,
<x_{n},y_{\sigma(n)}>_{{\cal H}_{1}})\eqno(3.20)
$$
if $B$ is symmetric, and:
$$
<B(x_{1},...,x_{n}),B(y_{1},...,y_{n})>_{{\cal H}_{0}} =
$$
$$
{1\over n!} \delta
\sum_{\sigma \in {\cal P}_{n}} (-1)^{\vert\sigma\vert}
\varphi(<x_{1},y_{\sigma(1)}>_{{\cal H}_{1}},...,
<x_{n},y_{\sigma(n)}>_{{\cal H}_{1}})\eqno(3.21)
$$
if $B$ is antisymmetric. In (3.19)-(3.20),
$\delta \in D_{0} \backslash \{ 0\}$.
\vskip0,5 truecm

{\bf Remark 16:} Let us supplement Definition 2 with:

(f) if
$a,b \in {\cal L}_{1}$
are in the center of
${\cal L}_{1}$,
then
$h(a,b) \in {\cal L}_{0}$
is in the center of
${\cal L}_{0}$

Then one can extend the results above to the case when
${\cal L}_{0}$
and
${\cal L}_{1}$
are propositional systems with superselection rules (see Remark 9).
\vskip1 truecm

B. It is interesting to see what gives our analysis in the case
when the one-particle system is an anyon i.e. a projective
unitary irreducible representation of the Poincar\'e group in $1
+ 2$ dimensions. It is tempting to see if one can recover the
multi-anyonic wave function.

For the identification of the one-particle system we rely on the
results of [13] where the complete list of projective unitary
irreducible representations of the Poincar\'e group in $1 + 2$
dimensions is given.

We provide here the necessary information. One should consider
the unitary irreducible representations of the universal
covering group of the Poincar\'e group in $1 + 2$ dimensions.

First, one identifies the universal covering group of the
Lorentz group in $1 + 2$ dimensions with:
$$
G \equiv R \times \{ u \in C \vert~~ \vert u\vert < 1 \} \eqno(3.22)
$$
with the composition law:
$$
(x,u)\cdot (y,v) \equiv \left( x+y+{1\over 2i} ln
{1+e^{-2iy}u\bar{v} \over 1+e^{2iy}v\bar{u}}, {u+e^{2iy}v \over
e^{2iy}+u\bar{v}} \right).\eqno(3.23)
$$
(one writes
$
{1 + z \over 1 + \bar{z}} \in C_{1} -\{-1\}
$
uniquely as
$
e^{2it}
$
with
$
t \in \left( -\pi /2, \pi/2 \right)
$).

Next, one provides the covering homomorphism
$\delta: G \rightarrow {\cal L}^{\uparrow}_{+}$
by composing
$
\delta_{1}: G \rightarrow SL(2,R)
$
given by:
$$
\delta_{1}(x,u) \equiv {1 \over 2\sqrt{1 - \vert u\vert^{2}}}
\left( \matrix {e^{ix}(1+u) + e^{-ix}(1+\bar{u}) &
ie^{ix}(1-u) - ie^{-ix}(1-\bar{u}) \cr
-ie^{ix}(1+u) + ie^{-ix}(1+\bar{u})  &
ie^{ix}(1-u) + ie^{-ix}(1-\bar{u}) } \right)\eqno(3.24)
$$
with
$\delta_{2}: SL(2,R) \rightarrow {\cal L}^{\uparrow}_{+}$
constructed in analogy to the covering map
$SL(2,C) \rightarrow {\cal L}^{\uparrow}_{+}$
(in $1 + 3$
dimensions).

Finally, one identifies the universal covering group of the
Poincar\'e group in $1 + 2$ dimensions with the inhomogeneous
group associated to $G$:
$$
in(G) = G \times R^{3}\eqno(3.25)
$$
with the composition law:
$$
((x,u),a) \circ ((y,v),b) = ((x,u) \circ (y,v),a + \delta (x,u)
b).\eqno(3.26)
$$

The analysis [13] provides a complete list of all unitary
irreducible representations of
$in(G)$.
We give only the formulae for the systems of non-zero mass and
of zero mass, leaving aside the tachions and the representations
of null momentum. We denote by
$X_{m}^{\pm}$
the upper (lower) hyperboloid of mass
$m \in R_{+} \cup \{ 0\}$
and by
$\alpha_{m}^{\pm}$
the corresponding Lorentz invariant measures. Then, the particle
of non-zero mass are identified to the representation
$W^{m,\eta,s}~~~(m \in R_{+}, \eta = \pm, s \in R)$
acting in
$L^{2}(X_{m}^{\eta},d\alpha^{\eta}_{m})$
as follows:
$$
\left( W^{m,\eta,s}_{x,u,a}f\right)(p) = e^{i\{a,p\}} e^{isx}
\left[ {p^{0}+\eta
m-\bar{u}e^{2ix}<p> \over p^{0}+\eta
m-ue^{-2ix}\bar{<p>}}\right]^{s/2} f(\delta(x,u)^{-1} p).\eqno(3.27)
$$

The particles of zero mass are identified to the representations
$W^{\eta,s,t}~~~(\eta = \pm, s \in R(mod~~2), t \in R)$
acting in
$L^{2}(X_{0}^{\eta},d\alpha^{\eta}_{0})$
as follows:
$$
\left( W^{\eta,s,t}_{x,u,a}f\right)(p) = e^{i\{a,p\}} e^{isx}
\left[ {p^{0}
-\bar{u}e^{2ix}<p> \over p^{0}
-ue^{-2ix}\bar{<p>}}\right]^{s/2}\times
$$
$$
exp\left\{i\eta t{Im(ue^{-2ix}\bar{<p>}) \over
p^{0}[(1+\vert u\vert^{2}) p^{0}-
2Re(ue^{-2ix}\bar{<p>})}\right\}
f(\delta(x,u)^{-1} p).\eqno(3.28)
$$

Here
$<p> \equiv p^{1} + i p^{2}$
and
$\{a,p\} \equiv a^{0}p^{0} - a^{1}p^{1} - a^{2}p^{2}$

We will take in the scheme of A as the one-particle space
${\cal H}_{1}$
one of the two possibilities above. Next, we should decide about
the statistics: Bose or Fermi?. To discriminate between these
two possibilities we procceed as follows. We apply the standard
construction of the field operator for both statistics and check
the causality. In the Bose case the commutator
$[\phi(x),\phi(y)]$
should vanish for
$x - y$
a space-like vector and in the Fermi case the same should happen
for the anti-commutator
$\{[\phi(x),\phi(y)\}$.
Simple computations show (see e.g. [14], ch. 3) that in the
first case the commutator is proportional to the antisymmetric
Pauli-Jordan distribution and in the second case the
anti-commutator is proportional to the symmetric Pauli-Jordan
distribution. Only the first case verifies the causality
condition required above. So, we are entitled to conclude that
we should choose Bose statistics. This statement is a very
primitive spin-statistics type theorem.

Now we restrict (3.27) or (3.28) to the universal covering group
of the Euclidean group in $1 + 2$ dimensions, i.e. we put
$u = 0$
and
$a = (0,{\bf a})$.
We get in both cases a representation which can be realized in
${\cal H}_{1} = L^{2}(R^{2},d {\bf p})$
and acts as follows:
$$
\left(U_{x,{\bf a}} f\right)({\bf p}) = e^{-i{\bf a}\cdot {\bf p}}
e^{isx} f\left( R(x)^{-1} {\bf p}\right).\eqno(3.29)
$$

Here:
$$
R(x) = \left( \matrix {cos(x) & -sin(x) \cr sin(x) & cos(x)}
\right).\eqno(3.30)
$$

For the composite system of $n$ such subsystems we use Bose
statistics (as justified above) and obtain the Hilbert space:
$$
\lor^{n} {\cal H}_{1} \cong \{ f:R^{\times n} \rightarrow C \vert
\int \vert f\vert^{2} d {\bf p}_{1}... d {\bf p}_{n} < \infty,
$$
$$
f({\bf p}_{\sigma(1)},...,{\bf p}_{\sigma(n)}) =
f({\bf p}_{1},...,{\bf p}_{n}),~\forall \sigma \in {\cal P}_{n}\}
$$
and the representation
$$
\left(U^{\otimes n}_{x,{\bf a}} f\right) ({\bf p}_{1},...,{\bf
p}_{n}) = e^{-i {\bf a}\cdot({\bf p}_{1} + ... + {\bf p}_{n})}
e^{isnx} f(R(x)^{-1} {\bf p}_{1},...,R(x)^{-1} {\bf p}_{n}).\eqno(3.31)
$$

Next, we perform a Fourier transform and end up with a
representation acting in the Hilbert space:
$$
{\cal H}_{0} = \{ f:(R^{2})^{\times n} \rightarrow C \vert
\int \vert f\vert^{2} d {\bf x}_{1}... d {\bf x}_{n} < \infty,
f({\bf x}_{\sigma(1)},...,{\bf x}_{\sigma(n)}) =
f({\bf x}_{1},...,{\bf x}_{n}),~\forall \sigma \in {\cal P}_{n}\}
$$
according to the formula:
$$
\left(U^{\otimes n}_{x,{\bf a}} f\right) ({\bf x}_{1},...,{\bf
x}_{n}) = e^{isnx}
f(R(x)^{-1} ({\bf x}_{1} - {\bf a}),...,
R(x)^{-1} ({\bf x}_{n} - {\bf a})).\eqno(3.32)
$$

{\bf Remark 17:} We note that this formula indicates that the
total spin of the system is $ns$.
\vskip0.5 truecm

{\bf Remark 18:} Let us define the configuration space:
$$Q_{n} \equiv ((R^{2})^{\times n} \backslash C) /
{\cal P}_{n}.\eqno(3.33)
$$

Here $C$ is the so-called {\it collision set}:
$$
C \equiv \{ ({\bf x}_{1},...,{\bf x}_{n}) \in (R^{2})^{\times n}
\vert {\bf x}_{i} = {\bf x}_{j}~for some~~i \not= j\}\eqno(3.34)
$$
and one factorizes to the natural action of
${\cal P}_{n}$.

Then, it is clear from (3.32) that the composite system we have
described is localizable, in the sense of Newton-Wigner-Wightman
[2] on the configuration space
$Q_{n}$
i.e. one can work in
${\cal H}_{0} = L^{2}(Q_{n})$.

It is more convenient to identify
$R^{2}$
with the complex plane $C$ as follows:
$z_{j} = x_{j}^{1} + i x_{j}^{2}~(j = 1,...,n),~\alpha = a^{1} +i
a^{2}$. In these new variables the Hilbert space is:
$$
{\cal H}_{0} = \{f:C^{\times n} \rightarrow C \vert \int \vert
f\vert^{2} dz_{1} d\bar{z}_{1}...dz_{n} d\bar{z}_{n} < \infty,
f(z_{\sigma(1)},...,z_{\sigma(n)}) = f(z_{1},...,z_{n})\}.\eqno(3.35)
$$
The formula for the representation (3.32) takes the form:
$$
\left(U^{\otimes n}_{x,{\bf a}} f\right) (z_{1},...,z_{n}) = e^{isnx}
f(e^{-ix} (z_{1} - \alpha),...,e^{-ix} (z_{n} - \alpha)).\eqno(3.36)
$$

Now we proceed as follows [15]. To every element
$f \in {\cal H}_{0}$
we associate the multiform function $F$ defined on the universal
covering space of
$Q_{n}$:
$$
F(z_{1},...,z_{n}) = \prod_{j<k} \left( z_{j} -
z_{k}\right)^{2\theta} f(z_{1},...,z_{n})\eqno(3.37)
$$
(see [15], eq. (1.15)).

In this new representation (3.36) becomes:
$$
\left(U^{\otimes n}_{x,{\bf a}} F\right) (z_{1},...,z_{n}) =
e^{i(sn+\theta n(n-1))x}
F(e^{-ix} (z_{1} - \alpha),...,
e^{-ix} (z_{n} - \alpha)).\eqno(3.38)
$$

Now we concentrate on the expression of the total Hamiltonian.
It is clear that in the representation (3.35)+(3.36) the total
Hamiltonian is the sum of the free one-particle Hamiltonians:
$$
H = \sum_{j=1}^{n} H_{0}(\nabla_{j}).\eqno(3.39)
$$

Let us try however like in [15], eq.(1.17) to consider that the
total Hamiltonian is the sum of free one-particle Hamiltonians
in the new representation (3.37). In this case, reverting to the
old representation, one obtains a topological interaction, i.e.
the total Hamiltonian is:
$$
H = \sum_{j=1}^{n} H_{0}(\nabla_{j} - i A(z_{j}))\eqno(3.40)
$$
where:
$$
A_{\mu}(z_{j}) = - \theta \sum_{k \not= j} \epsilon_{\mu\nu}
{x_{j}^{\nu} - x_{k}^{\nu} \over \vert z_{j} -
z_{k}\vert^{2}}.\eqno(3.41)
$$

We have obtained the so-called system of $n$ "free" anyons, as
presented in [16]-[21]. So, we can conclude, like in [15] that a
system of "free" anyons is equivalent to a system of bosons
carrying a charge $e$ and a magnetic vorticity $\Phi$ with:
$e\Phi = - \theta$.

{\bf Remark 19:} One notes from (3.37) that by exchanging two
variables , say:
$z_{j} \leftrightarrow z_{k}$,
one gets a sign
$(-1)^{2\theta}$.
So, it is tempting to call $\theta$ the "statistics" of the
system and say that one has interpolating statistics between
Bose and Fermi statistics. However, we think that it is more
natural to stick to the interpretation above, namely to conclude
that a system of anyons is nothing else but a system of bosons
with a special topological interaction.
\vskip0.5 truecm

{\bf Remark 20:} One may wonder if there exists a connection
between $\theta$ and $s$. An analysis based on the algebraic
framework of quantum theory gives:
$s = \theta~(mod~1)$
[6], [9]. Probably the same conclusion can be derived in a more
simpler way, using, as above, the argument of causality with the modified
Hamiltonian (3.40).
\vskip1 truecm

C. Finally, we try to explain why one does not obtain in our
analysis exotic statistics as parastatistics and infinite
statistics [6] which appear naturally in the algebraic
formulation of quantum theory. For this, one has to make a
comparison between the algebraic and the lattice-theoretical
formulation of quantum theory.

The first question to settle is: what is the algebra of
observables for a pure quantum system
${\cal L} = {\cal L}({\cal H})$?
According to [1] (see also Subsection 2A), every orthogonal
projector in
${\cal H}$
is an (elementary) observable of the sytem. This implies that
the algebra of observables of
${\cal L}({\cal H})$
is
${\cal B}({\cal H})$
i.e. the set of all bounded self-adjoint operators in
${\cal H}$.
That's it, the algebra of observables is the "largest" possible
one. But comparing to the analysis of [6] we note that in this
case we are left only with Bose and Fermi statistics. In fact,
exotic statistics can appear only if the algebra of observables
is "smaller". Even in this case one can prove that any system
with parastatistics can be converted into a system with normal
statistics (Bose or Fermi) by enlarging the algebra of
observables: namely, one can prove the existence of a gauge
group of symmetry $G$ and then a system with parastatistics can be
transformed into a system with normal statistics but living in a
non-trivial representation of the gauge group $G$ [22].

So, there seems to be a physical agreement between our result
and the corresponding algebraic analysis. We might note however
that our result is quite independent of the space-time
localization properties of the physicsl system. On the contrary,
such properties play a major r\^ole in the algebraic framework.
\vskip 1truecm
{\bf 4. Conclusions}

We have succeeded to prove that the lattice-theoretical (or
propositional) point of view on quantum physics is compatible,
under very general assumptions, only with two kind of statistics:
Bose-Einstein and Fermi-Dirac. We have also succeeded to show
that there exists some "philosophical" agreement between our
result and the similar analysis appearing in the litterature,
in the algebraic framework.

As we have explained in Subsection 3D, the game seeems to be
more simpler than in the algebraic formalism because the algebra
of observables is "too large". In fact, for the same reason, all
automorphisms of the algebra of observables are unitary (or
antiunitary) implementable in the lattice-theoretical framework,
so we cannot describe the phenomenon of spontaneous breakdown of
symmetry. (This phenomenon appears when the algebra of
observables admits automorphisms which are not unitary
implementable).

So, an interesting direction is suggested. Namely, one should
try to generalize somehow the lattice-thoretical framework such
that the corresponding algebra of observables is strictly
smaller than
${\cal B}({\cal H})$.
If this can be accomplished, then it is plausible that more
phenomenae as the spontaneous breakdown of symmetry could be
accomodated in this framework.
\vskip 1truecm
\vfil\eject

{\bf References}

\item{1.} C. Piron, "Foundation of Quantum Physics", W. A.
Benjamin, 1976
\item{2.}
V. S. Varadarajan, ``Geometry of Quantum Theory'' (second edition),
Springer, 1985
\item{3.}
D. Aerts, I. Daubechies, "Physical Justification for Using the
Tensor Product to Describe two Quantum Systems as one Joint
System", Helv. Phys. Acta 51 (1978) 661-675
\item{4.}
D. Aerts, "Subsystems in Physics Described by Bilinear Maps
Between the Corresponding Vector Spaces", Journ. Math. Phys. 21
(1980) 778-788
\item{5.}
D. Aerts, "Description of Compound Physical Systems and Logical
Interaction of Physical Systems", in "Current Issues in Quantum
Logic" (E. G. Beltrametti, B. C. van Frassen eds.), Plenum, 1981
\item{6.}
S. Doplicher, R. Haag, J. E. Roberts, "Local Observables and
Particle Statistics I", Commun. Math. Phys. 23 (1971) 199-230
\item{7.}
U. Uhlhorn, "Representations of Symmetry Transformations in
Quantum Mechanics", Arkiv. f. Fysik 23 (1962) 307-340
\item{8.}
J. Fr\"ohlich, P. A. Marchetti, "Quantum Field Theories of Vortices
and Anyons", Commun. Math. Phys. 121 (1989) 177-223
\item{9.}
J. Fr\"ohlich, F. Gabiani, "Braid Statistics in Local Quantum
Theory", Review in Math. Phys. 2 (1990) 251-353
\item{10.}
D. Finkelstein, J. M. Jauch, S. Schiminovich, D. Speiser,
"Foundation of Quaternionic Quantum Mechanics", Journ. Math.
Phys. 3 (1962) 207-220
\item{11.}
W. Greub, "Multilinear Algebra" (second edition), Springer, 1978
\item{12.}
D. Aerts, "Construction of a Structure which Enables to Describe
the Joint System of a Classical System and a Quantum System",
Rep. Math. Phys. 20 (1984) 117-129
\item{13.}
D. R. Grigore, "The Projective Unitary Irreducible
Representations of the Poincar\'e Group in $1 + 2$ Dimensions",
hep-th/9304142, submitted to Journ. Math. Phys.
\item{14.}
C. Itzykson, J. B. Zuber, "Quantum Field Theory", McGraw-Hill, 1980
\item{15.}
J. Fr\"ohlich, "Quantum Statistics and Locality", in the
Proceedings of the Gibbs Symposium, Yale, 1989 (D. G. Caldi, G.
D. Mostow eds.)
\item{16.}
M. G. G. Laidlaw, Cecile Morette DeWitt, "Feynman Functional
Integrals for Systems of Indistinguishable Particles", Phys.
Rev. D 3 (1971) 1375-1378
\item{17.}
J. M. Leinaas, J. Myrheim, "On the Theory of Identical
Particles", Il Nuovo Cimento 37B (1977) 1-23
\item{18.}
G. A. Goldin, R. Meinkoff, D. H. Sharp, "Representation of the
Local Current Algebra in a Non-Simply Connected Spaces and the
Aharonov-Bohm Effect", Journ. Math. Phys. 22 (1981) 1664-1668
\item{19.}
Yong-Shi Wu, "General Theory for Quantum Statistics in Two
Dimensions", Phys. Rev. Lett. 52 (1984) 2103-2106
\item{20.}
A. P. Balachandran, "Classical Topology and Quantum Statistics",
in "Fractional Statistics and Anyon Superconductivity" (F.
Wilczek ed.), World Scientific, 1991
\item{21.}
S. Forte, "Quantum Mechanics and Field Theory with Fractional
Spin and Statistics", Rev. Mod. Phys. 64 (1992) 193-236
\item{22.}
S. Doplicher, J. E. Roberts, "Why there is a Field Algebra with
a Compact Gauge Group Describing the Superselection Structure of
Particle Physics", Commun. Math. Phys. 131 (1990) 51-107
\bye